\documentclass[larticle]{aa}
\usepackage{natbib}
\usepackage{graphicx}
\usepackage{times}
\usepackage{amsmath}
\usepackage{txfonts}
\usepackage{color}
\usepackage{ulem}
\newcommand{\etal}{et al.,~}
\newcommand{\msun}{{\,\rm M}_{\odot}}
\newcommand{\lsun}{{\,\rm L}_{\odot}}

\newcommand{\hh}{\ifmmode {{\rm H}_2} \else {H$_2$} \fi}
\newcommand{\nht}{\ifmmode {{\rm NH}_3} \else {NH{\bas 3}} \fi}
\newcommand{\tco}{\ifmmode {^{13}{\rm CO}} \else {$^{13}{\rm CO}$}\fi}
\newcommand{\dco}{\ifmmode {^{12}{\rm CO}} \else {$^{12}{\rm CO}$}\fi}
\newcommand{\cdo}{\ifmmode {{\rm C}^{18}{\rm O}} \else {${\rm C}^{18}{\rm O}$}\fi}

\newcommand{\htco}{\ifmmode {{\rm H}^{13}{\rm CO}^{+} } \else {${\rm H}^{13}
{\rm CO}^{+}$ }\fi}
\newcommand{\hco}{\ifmmode {{\rm H}^{12}{\rm CO}^{+} } \else {${\rm H}^{12}
{\rm CO}^{+}$ }\fi}
\newcommand{\juz}{\ifmmode {{\rm J}=1\rightarrow 0} \else
{J=1$\rightarrow$0}\fi}
\newcommand{\jdu}{\ifmmode {{\rm J}=2\rightarrow 1} \else
{J=2$\rightarrow$1}\fi}
\newcommand{\jtd}{\ifmmode {{\rm J}=3\rightarrow 2} \else
{J=3$\rightarrow$2} \fi}
\newcommand{\jcq}{\ifmmode {{\rm J}=5\!\rightarrow\!4} \else
{${\rm J}=5\!\rightarrow\!4$} \fi}
\newcommand{\as}{\ifmmode {^{\scriptscriptstyle\prime\prime}}
        \else $^{\scriptscriptstyle\prime\prime}$\fi}
\newcommand{\am}{\ifmmode {^{\scriptscriptstyle\prime}}
        \else $^{\scriptscriptstyle\prime}$\fi}
\renewcommand{\hco}{\ifmmode {{\rm HCO}^+} \else {HCO$^+$} \fi}

\begin{document}
\title{CID: Chemistry in disks \\
VI.sulfur-bearing molecules in the protoplanetary disks surrounding LkCa15, MWC480, DM Tau, and GO Tau
\thanks{Based on observations carried out with the IRAM
30m radiotelescope.
IRAM is supported by INSU/CNRS (France), MPG (Germany), and IGN
(Spain).}}

\author{Anne Dutrey\inst{1,2}, Valentine Wakelam\inst{1,2}, Yann Boehler\inst{1,2}, St\'ephane Guilloteau \inst{1,2}, Franck Hersant\inst{1,2},
Dmitry Semenov \inst{3}, Edwige Chapillon\inst{4}, Thomas Henning \inst{3},  Vincent Pi\'etu \inst{5}, Ralf Launhardt \inst{3}, Frederic Gueth
\inst{5}, Katharina Schreyer \inst{6}}
\offprints{A.Dutrey \email{dutrey@obs.u-bordeaux1.fr}}
\institute{Universit\'e de Bordeaux, Observatoire Aquitain des Sciences de l'Univers (OASU), 2 rue de l'Observatoire,
BP89, F-33271 Floirac Cedex, France
\and{}
CNRS - UMR5804, Laboratoire d'Astrophysique de Bordeaux (LAB), 2 rue de l'Observatoire, BP 89, F-33271 Floirac Cedex, France
\and{}
Max-Planck-Institut f\"ur Astronomie, K\"onigstuhl 17, D-69117
Heidelberg, Germany
\and{}
MPIfR, Auf dem H\"ugel 69, 53121 Bonn, Germany.
\and{} IRAM, 300 rue de la piscine, F-38406
Saint Martin d'H\`eres, France
\and{} Astrophysikalisches Institut und Universit\"ats-Sternwarte,
Schillerg\"asschen 2-3, D-07745
Jena, Germany }
\date{Received 21-mar-2011, Accepted 29-July-2011}
\authorrunning{Dutrey et al \etal}
\titlerunning{Sulfur-bearing molecules in protoplanetary disks}

  \abstract
{}
{We study the content in S-bearing molecules of protoplanetary disks around low-mass stars.}
{We used the new IRAM 30-m receiver EMIR to perform simultaneous observations of the $1_{10}-1_{01}$ line of H$_2$S at 168.8 GHz and
$2_{23}-1_{12}$ line of SO at 99.3 GHz. We compared the observational results with predictions coming from the astrochemical code
NAUTILUS, which has been adapted to protoplanetary disks. The data were analyzed together with existing CS J=3-2 observations.}
{We fail to detect the SO and H$_2$S lines, although CS is detected in LkCa15, DM\,Tau, and GO\,Tau but not in MWC\,480.
However, our new upper limits are significantly better than previous ones and allow us
to put some interesting constraints on the sulfur chemistry. }
{Our best modeling of disks is obtained for a C/O ratio of 1.2, starting from initial cloud conditions of H density of
$2\times 10^5$~cm$^{-3}$ and age of $10^6$~yr. The results agree with the CS data and are compatible with the SO upper limits, but fail
to reproduce the H$_2$S upper limits. The predicted H$_2$S column densities are too high by at least one order of magnitude.
H$_2$S may remain locked onto grain surfaces and react with other species, thereby preventing the desorption of H$_2$S.}


\maketitle{}

\section{Introduction}

Understanding the evolution of gas and dust accretion disks around young stars is one of the prerequisites
for handling the processes leading to planet formation. The temperature structure of disks is roughly
understood \citep{Chiang_etal1997,Menshchikov+1997,Dalessio_etal1999},
but very fundamental disk properties, such as the ionization degree and the dust and gas mass
distribution, are poorly constrained. As no direct tracer exists, the derivation of the gas mass distribution
requires in-depth knowledge of the disk structure and chemistry. Therefore, studying the chemistry becomes
a necessary step, because molecular abundances and gas distribution are strongly coupled in both
the modeling and analysis of observational data.

Vertically, the current paradigm of the chemistry of protoplanetary disks is
a layered model \citep[see for example][]{Zadelhoff_etal2001}. In the upper layer,
the large incident UV flux from the central object results in a photodissociation-dominated
layer (PDR) with a chemistry in equilibrium. The PDR size evolves with grain growth, since UV
flux can penetrate deeper into the disk \citep{Aikawa_Nomura_2006, Chapillon_etal2008, Vasyunin+2011}.
Below the PDR, the molecules are expected to concentrate in a warm molecular layer.
All chemical models published so far have produced similar results \citep{Aikawa_Nomura_2006,
Semenov_etal2005}.
Finally, because of the large extinction, the outer disk midplane is cold  ($\sim$ 10 K)
and molecules are expected to stick onto dust grains. The chemistry is then rather similar
to what is encountered in cold dense cores. Radially the ``snowline'', which is defined by the radius where
H$_2$O molecules start to evaporate from the ice mantle of dust grains, delineates
the zone where rocky planets can be formed. For a TTauri star similar to the young Sun,
the ``snowline'' at midplane should typically be located around $\sim 0.5-2$~AU.

The ideal scenario presented above is loosely constrained. There are already several
observational facts that reveal deficits in current chemical models for the outer disk that
contains the reservoir of gas and dust mass ($\sim$ R$> 30$~AU).
Very recently, sensitive observations with Herschel/HIFI have revealed that the emission of
H$_2$O is significantly weaker than predicted \citep{bergin+etal2010}.
Another example is given by our mm observations, which show ``cold'' molecules (at temperature
lower than $\sim$ 10 K) such as CO \citep{Dartois_etal2003,Pietu_etal2007},
C$_2$H \citep{henning+etal2010}, or CN and HCN \citep{Chapillon+etal2011}.
In the last cases, the disk density exceeds the critical density of the
observed transitions and subthermal excitation cannot be invoked \citep{Pavlyuchenkov_etal2007}.
The usual explanation given is the role of turbulence. In theory, the
vertical and radial mixing in the disk should allow a partial replenishment of the
cold midplane layer \citep{Semenov_etal2006}. However, this does not appear to be
sufficient by itself \citep{Hersant_etal2009}. The turbulent mixing efficiency
actually depends on the photodesorption rates \citep{Hersant_etal2009} which
 may be much higher than originally thought \citep{Oberg+etal2009}.

For inner disks, where planets should form ($\sim$ R$< 30$~AU), some very interesting
unresolved observations (mostly coming from the Spitzer satellite) begin to unveil the chemical
content of planet-forming regions. These observations show that not only CO but
also many other molecules such as H$_2$O, HCN, CO$_2$, C$_2$H$_2$, OH, etc\ldots are present
in inner disks \citep{Carr+etal2008}. So far, there has been no H$_2$S detection.
Interestingly, a recent paper from \citet{Glassgold+etal2009} suggests that there
is no need for transport in the inner nebula to explain the presence of water vapor at the
surface of the very inner disk. The disk irradiation by UV and X-rays may be sufficient to
create water ``in situ'', in the warm disk atmosphere. This model accounts roughly for the IR
water vapor emission observed in TTauri disks by \citet{Carr+etal2008} and \citet{Salyk+etal2008}
without radial or vertical mixing in disks.

Recent models of solar nebula investigating the sulfur chemistry \citep{Pasek+etal2005}
within the 20 central AU found that the main gas component of sulfur-bearing molecules would be
H$_2$S, while it should be in the form of FeS, MgS, or CaS in solid state (meteorites).
The abundance of those species are expected to be variable with respect to the time
evolution of  the snowline and the water vapor content.
Moreover, sulfur-bearing molecules such as H$_2$S and CS (and SO to a lesser extent) are
relatively abundant in comets. For example, in Hale-Bopp, the abundance of H$_2$S is
10\% that of CO. H$_2$S is supposed to come from the evaporation of ices, while CS and
SO are the daughters of CS$_2$ and SO$_2$, respectively \citep{Bockelee+etal2004}.
Nonetheless, the only S-bearing molecule that is firmly detected in cold outer TTauri disks is CS.

Several transitions of CS, the most abundant S-bearing species in cold molecular cores
\citep{Dickens+2000}, were detected by \citet{Dutrey_etal1997}, who found an abundance
($\sim 3 \cdot 10^{-10}$), about 30 times lower than in TMC1. Such a difference is
not surprising since the gas and dust have evolved inside the disk compared to dense cores.

In this paper, we go one step further by studying other S-bearing  molecules using
the IRAM 30-m radiotelescope.
The observations and results are described in S.2, the chemical modeling is presented in S.3, and we
discuss in S.4 the implications of our non detections before we conclude in S.5.

\section{Observations and results}

\begin{table*}
\caption{Sample of stars and stellar properties.}
\begin{center}
\begin{tabular}{lllllllll}
\hline \hline
Source   & Right Ascension         & Declination      & Spect.Type & Effective Temp. & Stellar Lum. & Stellar Mass & Age   & UV flux\\
         & ($^{o}$,$^{'}$,$^{''}$), (J2000.0) & ($^h$,$^m$,$^s$) (J2000.0) &            & (K)             & ($\lsun$)    & ($\msun$)    & (Myr) & ($\chi_0$) \\
\hline
LkCa~15  & 04:39:17.78 & 22:21:03.34  & K5 & 4350 & 0.74 & $1.01\pm0.02$ & 3-5 & 2550\\
DM~Tau   & 04:33:48.73 & 18:10:09.89  & M1 & 3720 & 0.25 & $0.53\pm0.03$ & 5   &  410\\
MWC~480  & 04:58:46.26 & 29:50:36.87  & A4 & 8460 & 11.5 & $1.83\pm0.05$ & 7   & 8500\\
GO~Tau & 04:43:03.05 & 25:20:18.8 & M0 & 3850 & 0.37 & 0.5  & 3  &   410 \\
 \hline
\end{tabular}
\end{center}
\tablefoot{Coordinates J~2000.0 deduced from the fit of the 1.3mm
continuum map of the PdBI performed in \citet{Pietu_etal2006}. Columns 3,4, 5, 6, and 7 are the spectral type,
effective temperature, stellar luminosity, and age as given in \citet{Simon_etal2000}. Masses are taken from
\citet{Pietu_etal2007}, for DM Tau, MWC480, and LkCa15. The stellar UV luminosities that are given in Col.
8 in units of the \citet{Draine1978} interstellar UV field are taken from \citet{Bergin+2004} (LkCa~15 and DM~Tau) or
computed from the \citet{Kurucz1993} ATLAS9 of stellar spectra (MWC~480 and GO Tau). They are given for a distance of
100 AU from the star. Half of this flux is supposed to scatter downwards inside the disk.} \label{tab:coord}
\end{table*}
\begin{figure*}
\begin{center}
\includegraphics[angle=270,width=16cm]{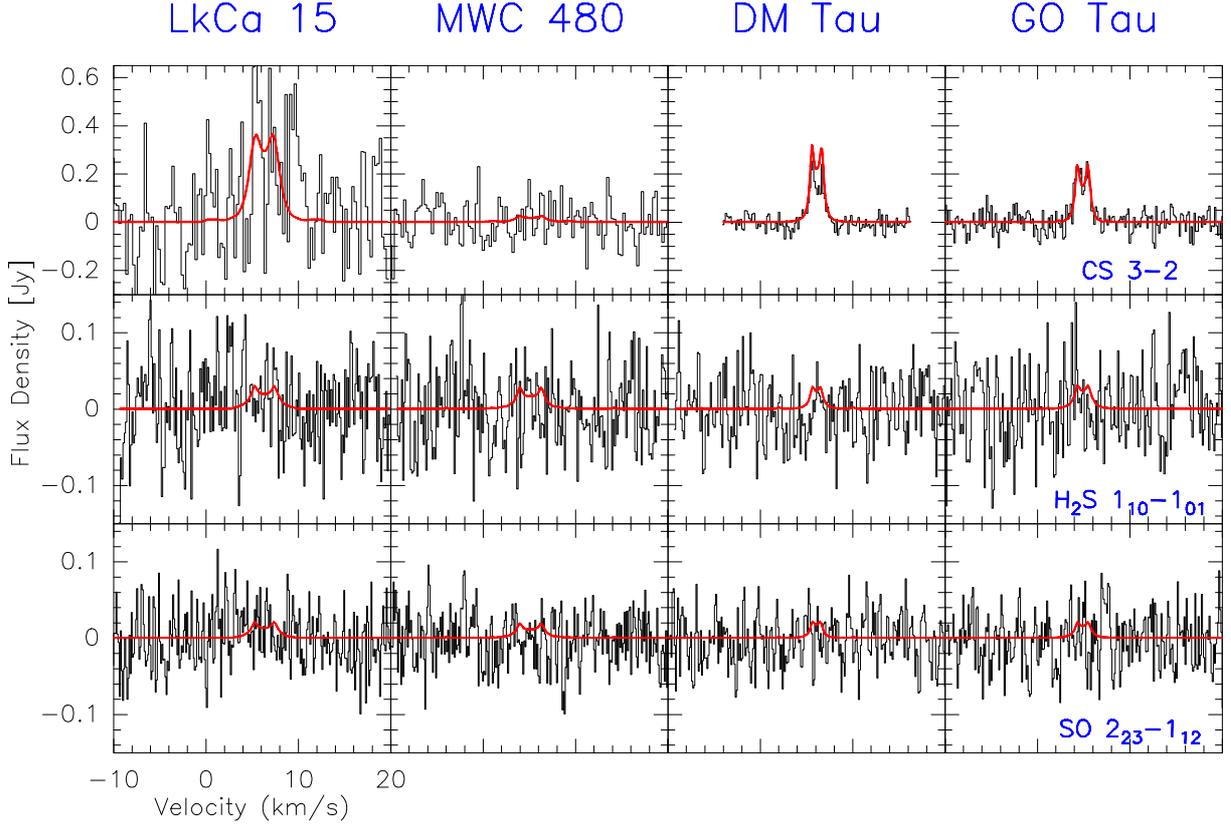}\\
\end{center}
\caption{Observations of SO $2_{23}-1_{12}$, H$_2$S $1_{10}-1_{01}$ and CS 3-2 in the four disks.
For CS J=3-2, the best models have been superimposed for all sources. For SO and H$_2$S,
the models always correspond to the 3$\sigma$ upper limits.}
\label{fig:spectre}
\end{figure*}

\subsection{Sample and observations}

 Many molecular lines have been observed
around MWC 480, LkCa15, and DM Tau \citep{Dutrey_etal1997}, and the disk structures are well known \citep{Pietu_etal2007}. DM Tau
and LkCa15 are TTauri stars of 0.5 and 1.0 $\msun$, respectively, while MWC480 is an Herbig Ae star of 1.8 $\msun$. The CO disk of GO Tau
is reported in \citet{schaeffer_etal2009}. It appears to be a good candidate
for a deep search of chemistry since it exhibits a large disk (CO outer radius R$_{out} \simeq 900$~AU) that is
very similar to that of DM Tau. Table \ref{tab:coord}
summarizes the stellar properties of the four stars.

The observations were performed using the new 30-m heterodyne receivers (EMIR) mid-May
2010 under good weather conditions. We used the wobbler switch mode with offsets of 60$''$
since the sources are unresolved in the telescope beam, providing spatially
integrated spectra. We observed the H$_2$S $1_{10}-1_{01}$ line at 168.763 GHz simultaneously with the SO $2_{23}-1_{12}$
line at 99.299 GHz using the correlator VESPA with a spectral resolution of 20 KHz. Venus
was regularly checked for pointing and focus corrections. The forward
efficiency was set to 0.95 and 0.93 at 99 and 168 GHz, respectively. We used flux density
conversion factors (Jy/T$_A^*$) of 6 Jy/K at 99 GHz and 6.5 Jy/K at 168 GHz.
We took the values recommended by the IRAM staff
(see http://www.iram.es/IRAMES/mainWiki/Iram30mEfficiencies).

The CS J=3-2 data of GO Tau were observed in 2009 at the IRAM 30-m
radiotelescope using EMIR and reduced as described above.
The CS J=3-2 data of MWC\,480 and LkCa15 were obtained with the IRAM 30-m telescope
in 2000 simultaneously with the DCO$^+$ J=2-1 observations published in \citet{Guilloteau_etal2006},
using the same observational strategy as described in \citet{Dutrey_etal1997}.
For DM Tau, CS J=3-2 was mapped with the IRAM array in the course of an imaging project (Guilloteau et al.,
2011, in prep.) and we only present here the integrated spectrum (integrated over the CO outer radius).

All the 30-m data were reduced using CLASS. The DM Tau CS map was analyzed using CLIC and MAPPING
and an integrated spectrum was produced.

\subsection{Column density derivation and results}

 Figure \ref{fig:spectre} shows the integrated spectra obtained
for all the sources. Neither H$_2$S nor SO are detected but the observations provide for the first
time meaningful upper limits, significantly better (by a factor $\sim 7-10$)
than those quoted in \citet{Dutrey+2000}.
CS J=3-2 is detected in three disks: GO Tau, LkCa15, and DM Tau, but not in MWC\,480.

\begin{table}
\caption{Physical parameters used to derive the best fit models and the surface
density upper limits}
\begin{tabular}{lcccc}
\hline\hline
Source        & MWC~480         &  LkCa~15 & DM~Tau & GO~Tau\\
\hline
inclination ($^{o}$) & 38 & 52 & -32& 51\\
P.A.($^{o}$) & 57 & 150 & 65 & 112 \\
\hline
V$_{syst}$(km.s$^{-1}$) &5.10 & 6.30& 6.04& 4.89 \\
V$_{100}$(km.s$^{-1}$) &4.03 &3.00 &2.16 & 2.05 \\
$\delta_v$(km.s$^{-1}$) &0.2 & 0.2 & 0.2 & 0.2 \\
\hline
 T$_{100}$(K)  & 30  & 15   & 15 & 15   \\
R$_{int}$(AU)       &   1 &  45 & 1 & 1 \\
R$_{out}$(AU)       &   500 & 550  & 800 & 900 \\
\hline
CS R$_{out}$(AU)       &   - & -  & 560$\pm$ 10 & - \\
CS $p$ &   - & -  & 0.75$\pm$0.05 & 1.0$\pm$0.2 \\
\hline
\end{tabular}
\tablefoot{P.A. is the position angle of the disk rotation axis, $i$ the inclination
angle of the disk (0 means face-on), V$_{syst}$ the systemic velocity.
The velocity laws (V$_{100}$) are Keplerian \citep{Pietu_etal2007,schaeffer_etal2009}.
$\delta_v$ is the turbulent line-width component.
The temperatures and surface densities follow the laws
$T(r) = T_{100}(r/100\mathrm{AU})^{-0.5}$ and
$\Sigma(r) = \Sigma_{300}(r/300 \mathrm{AU})^{-1.5}$, respectively.
For CS, in the case of DM Tau, the outer radius and the surface density
radial dependence $p$ derived from the data have been used. This is also the
case for $p$ in GO Tau. Derived parameters are given with their errorbars
(a ``$-$'' means that the standard values are assumed.)
\label{tab:param}}
\end{table}
Deriving proper estimates of upper limits is complicated
in the case of Keplerian disks because of the existence of velocity, temperature,
and density gradients \citep{Guilloteau_etal2006, Dutrey_etal2007}.
For chemistry, since the surface density and the
temperature have radial dependencies, defining an averaged column density over the whole disk
implies to merge chemically different regions. This can only be
properly done  by taking the existing gradients into account.
Because the characteristic size of the disks is several 100 AU,
and is unresolved in the 30-m beam, our observations are most sensitive
to the 200 - 400 AU range, depending on the radial distribution
\citep[see][]{Pietu_etal2007}.
We thus chose to proceed as follows. We used DISKFIT \citep[a dedicated radiative
transfer code to protoplanetary disks, see][]{Pietu_etal2007} to generate
integrated spectra and derived the best model by adjusting disk parameters.
The models provided by DISKFIT were compared to the observed spectra
using the minimization routine described in \citet{Pietu_etal2007}. For the
minimizations, all parameters were fixed except $\Sigma_{300}$, the surface
density at 300 AU. The adopted disk parameters are given in Table
\ref{tab:param}. Theses values come from the angularly resolved CO
interferometric analysis performed by \citet{Pietu_etal2007} and
for GO Tau by \citet{schaeffer_etal2009}. The 3$\sigma$ upper limits on the
surface densities at 300\,AU were then calculated from the formal error
obtained from the best fit. The results are given in Table \ref{tab:coldens}.
The choice of the CO parameters to fit sulfur-bearing molecules is governed
by the fact that CO results provide the best constraints
on the gas disk structure and kinetic temperature. Moreover, the temperature derived
from CO data can be considered as representative of a significant fraction of the
molecular layer so is well suited to determining upper limits on the surface density of molecular species.
Two parameters may affect $\Sigma_{300}$: the exponent slope $p$ and the outer radius R$_\mathrm{out}$.
$\Sigma_{300}$ itself is relatively unsensitive to these parameters, but
the derived error-bars can be significantly affected. On one hand, for DM Tau,
using $p = 0.5$ instead of 1.5 decreases the error by a factor 4. On the other
hand, further reducing R$_\mathrm{out}$ to 560 AU (the best fit value for CS in DM Tau)
 again  increases the error by a factor 2. Thus our choice of values for R$_\mathrm{out}$
and $p$ is rather conservative in view of the unknowns.
\begin{table}
\caption{Sulfur-bearing molecules: detections and 3$\sigma$ upper limits.}
\centering
\begin{tabular}{l||ccc}
\hline
\hline
& & & \\
 Sources & \multicolumn{3}{c}{$\Sigma_{300}$ (cm$^{-2}$)} \\
 & SO & H$_2$S & CS \\
\hline
 & & \\
DM Tau & $\leq$ 7.5$\cdot$10$^{11}$ & $\leq$ 1.4$\cdot$10$^{11}$ & 3.5 $\pm$ 0.1$\cdot$ 10$^{12}$\\
LkCa15 & $\leq$ 1.9$\cdot$10$^{12}$ & $\leq$ 3.6$\cdot$10$^{11}$ & 8.7 $\pm$ 1.6$\cdot$ 10$^{12}$\\
MWC480 & $\leq$ 2.5$\cdot$10$^{12}$ & $\leq$ 4.1$\cdot$10$^{11}$ & $\leq$ 8.4 $\cdot$10$^{11}$\\
GO Tau & $\leq$ 8.9 $\cdot$10$^{11}$ & $\leq$ 1.8 $\cdot$10$^{11}$ & 2.0 $\pm$ 0.16$\cdot$ 10$^{12}$ \\
\hline
\end{tabular}
\tablefoot{Surface densities at 300 AU (modeled as
$\Sigma(r)= \Sigma_{300}(r/300 \mathrm{AU})^{-1.5}$)
derived from the 30-m data (except for CS 3-2 in DM Tau)
and the model DISKFIT. See text for details.}
\label{tab:coldens}
\end{table}

\section{Chemical analysis and modeling}
%
\label{mod_desc}

\begin{figure}
\includegraphics[width=0.6\linewidth]{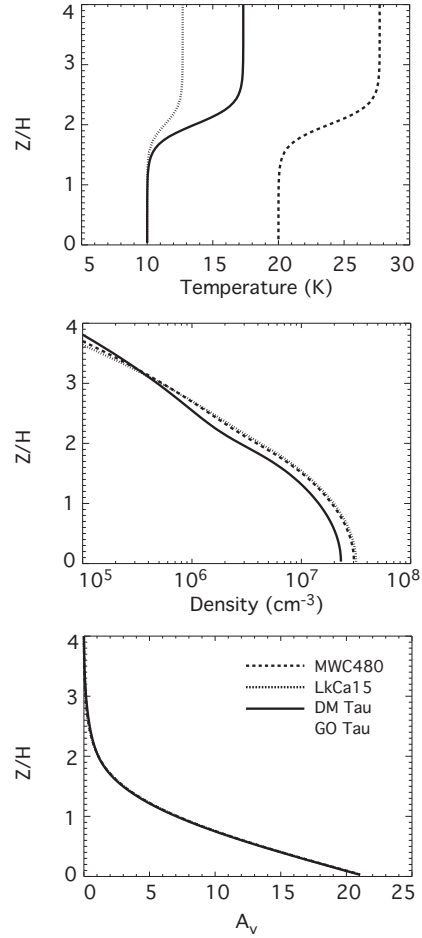}
\caption{Temperature, H$_2$ density, and visual extinction vertical profiles in the four protoplanetary
disks at 300~AU used for the chemical modeling. GO Tau and DM Tau have very similar physical parameters so that
we only show the curves for DM Tau.  \label{fig:phys_disk}}
\end{figure}

\begin{table}
\caption{Initial abundances of various species (/H) for the parent cloud.}
\begin{center}
\begin{tabular}{lccc}
\hline
\hline
Element & Abundance & Ref. & Case$^*$  \\
\hline
He & 0.09 & (1) & \\
N & 6.2(-5) & (2) &   \\
C$^+$ & 1.7(-4) & (2) & A,C \\
 & 1.2(-4) & (1) & B,D \\
O & 1.4(-4) & (3) & A,C \\
 & 2.56(-4) & (1) & B,D \\
S$^+$ & 8(-8) & (4) & A,B \\
 & 8(-9) & & C,D \\
Si$^+$ & 8(-9) & (4) & \\
Na$^+$ & 2(-9) & (4) &  \\
Fe$^+$ & 3(-9) & (4) &  \\
Cl$^+$ & 1(-9) & (4) &  \\
Mg$^+$ & 7(-9) & (4) & \\
P $^+$& 2(-10) & (4) &  \\
\hline
\end{tabular}
\end{center}
\tablefoot{References: (1) \citet{2008ApJ...680..371W}, (2)
\citet{2009ApJ...700.1299J}, (3) \citet{Hincelin+etal_2011}, and (4)
\citet{1982ApJS...48..321G}. $^*$ indicates
to which chemical model the corresponding values apply.}
\label{tab:elem_ab}
\end{table}
We used the NAUTILUS \citep{Hersant_etal2009} gas-grain model to study the sulfur chemistry in the four protoplanetary disks.
NAUTILUS computes the abundance of 460 gas-phase and 195 surface species as a function of time using the rate
equation method \citep{1992ApJS...82..167H}. The chemical network contains 4406 gas-phase reactions and 1733
reactions involving grains (including adsorption and desorption processes and grain-surface reactions).
The gas-phase network was updated according to the recommendations from the KIDA experts (http://kida.obs.u-bordeaux1.fr/, on November 2010).
The full network can be found at http://kida.obs.u-bordeaux1.fr/models/benchmark\_2010.dat. We used silicate grains of 0.1~$\mu$m size.
This small size agrees with the recent observational results from \citet{Guilloteau+2011}, which show that in
outer disks ($r>$70-100 AU) dust grains are dominated by small particles (ISM-like). This agrees with the
theoretical results of \citet{Birnstiel+etal_2010} and suggests that large grains have already migrated to
the inner disk regions ($r<$70 AU).
Since small grains of $0.1 \mu m$ should remain dynamically coupled to the gas, this observational result also
implies that the chemistry in the outer disk would be less affected by the dust settling than in the inner disk.
For the ratio of extinction curve to the column density of hydrogen, we used the standard
relation A$_\mathrm{V} = N(H) /1.6 \cdot 10^{21}$ following \citet{Wagenblast+1989}.
More details on the model can be found in \citet{Hersant_etal2009} and \citet{2010A&A...522A..42S}.
To obtain initial abundances for the chemistry of protoplanetary disks, we first compute the
chemical composition of the parent molecular cloud. With this aim,
we run NAUTILUS during $10^5$~yr for the following physical conditions: gas and dust temperature are 10~K, H
density is $2\times 10^4$~cm$^{-3}$ and A$_\mathrm{V}$ is 50 (models A and B).
We checked higher densities ($2\times 10^5$~cm$^{-3}$) and older clouds ($10^6$~yr), corresponding to models C and D
and also tested smaller A$_\mathrm{V}$ (10). For the cloud, the species (listed in Table 4) are assumed to initially be in the atomic form except for
hydrogen, which is entirely in H$_2$, and we assumed four sets of elemental abundances. In models A and C, we adopted the
oxygen elemental abundance required to reproduce the low O$_2$ abundance observed in the cold
ISM from \citet{Hincelin+etal_2011} and the carbon abundance from \citet{2009ApJ...700.1299J} for the $\zeta$Oph diffuse cloud,
so that the C/O elemental ratio is 1.2. In models B and D, we use the carbon and oxygen elemental abundances
from \citet[][$1.2\times 10^{-4}$ for C and $2.56\times 10^{-4}$ for O]{2008ApJ...680..371W}, which gives a
C/O ratio of 0.4. The choice of the sulfur elemental abundance ``available'' for the chemistry, i.e. not locked
into the refractory part of the grains, is a widely studied problem \citep{Ruffle+etal_1999,2004A&A...422..159W,
Scappini+etal_2003,vanderTak+etal_2003}. No depletion of gas phase atomic sulfur is observed in diffuse clouds compared
to the cosmic abundance of $10^{-5}$ (/H), in contrast to other elements. However, if such high elemental
abundance is used to do the chemistry in dense clouds, one would produce too many S-bearing molecules, on
which we have observational constraints \citep{Graedel+etal_1982,2008ApJ...680..371W}. For this reason,
depletion of sulfur in an unknown, more or less refractory form is assumed to happen very quickly when
the molecular cloud is formed from the diffuse gas. To simulate this effect, the ``free'' elemental abundance
of S is then decreased by an arbitrary amount. For simplicity, we have adopted the elemental abundance from
\citet{Graedel+etal_1982} in our Models A and B (which approximately reproduces the observations in dense clouds)
and decreased it by a factor of 10 in models C and D. Table~\ref{tab:model} also summarizes the
parameters (cloud age, initial density, C/O ratio) for the four Models A, B, C, and D. We did numerous
simulations to understand the effect of the parameters that we changed. In general, the S-bearing molecular
abundances are not very sensitive to these parameters, and we present here the extreme cases.

\begin{table}
\caption{Parameters used for the four models A, B, C, and D.}
\begin{center}
\begin{tabular}{lcccc}
\hline
\hline
Parameter & Model A  & Model B  & Model C & Model D  \\
\hline
C/O ratio & 1.2 & 0.4 & 1.2 & 0.4 \\
Sulfur abundance & $8 \times 10^{-8}$ & $8 \times 10^{-8}$& $8 \times 10^{-9}$ & $8 \times 10^{-9}$\\
 (/H) & & &  & \\
Cloud H density & $2 \times 10^{4}$ & $2 \times 10^{4}$& $2 \times 10^{5}$ & $2 \times 10^{5}$\\
 (H.cm$^{-3}$) & & & & \\
Cloud age & $ 10^{5}$ & $ 10^{5}$& $ 10^{6}$ & $ 10^{6}$\\
(yrs) & & & & \\
\hline
\end{tabular}
\end{center}
\label{tab:model}
\end{table}

The abundances of chemical species in the gas and on the grains were used as initial composition for the chemistry of
the disks, which is then integrated over $5\times 10^6$~yr, taken as the approximate age of the studied disks. The
computed column densities do not vary significantly after $10^6$~yr. For the disk physical parameters,
we used the two layer parametric disk model described in \citet{Hersant_etal2009}. Its temperature profile is based on \citet{Dartois_etal2003}.
The dust temperature is equal to the gas temperature.
The midplane is cold, with a temperature of $10$ K for DM Tau, GO Tau, and LkCa15 and $20$ K for MWC480, consistent with the temperature
derived from the $^{13}$CO 1-0 transition. The molecular surface layer is warmer, as derived from $^{12}$CO 2-1
observations, with a temperature of $17.3$ K for DM Tau and GO Tau, $12.7$ K for LkCa15, and $27.7$ K for MWC480 at 300 AU.
We assumed the same surface density for all four
disks, with a value of $0.15$ g cm$^{-2}$ at $300$ AU. The resulting disk structure is displayed Fig.
\ref{fig:phys_disk} \citep[see][for details about the computation of the physical structure]{Hersant_etal2009}.
For the UV radiative transfer in the disk, we consider only the vertical extinction (assuming that only half of UV photons are
scattered downwards by small grains located above our computing box).
The UV flux given at 100 AU in Table \ref{tab:coord} is decreasing as $1/r^2$ (where $r$ is the spherical radius) from the
central star. At a cylindrical radius of $300$ AU and a vertical boundary of four pressure
scale heights ($H$), the UV scale factor becomes $33$ for GO Tau, $17$ for DM Tau, $120$ for LkCa15, and $391$ for MWC480,
expressed in units of the InterStellar Radiation Field \citep[ISRF,][]{Draine1978}.

\begin{figure}
\includegraphics[width=1\linewidth]{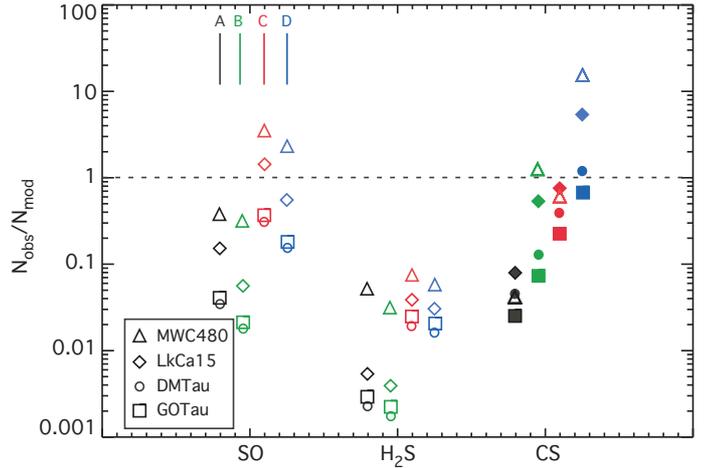}
\caption{Ratio between the column densities of SO, CS, and H$_2$S observed in the four protoplanetary disks at 300~AU from the central
star and the values computed with NAUTILUS. For SO and H$_2$S, they correspond to upper limits. The four Models (A, B, C, and D)
correspond to the elemental abundances given in Table \ref{tab:elem_ab} and to the parameters given in Table \ref{tab:model}.
 }
\label{fig:rap}
\end{figure}

Using the chemical and physical parameters, we ran the four Models A,B,C and D in order to obtain the column densities of SO,
H$_2$S, and CS at 300~AU. The results for Models A and C are given in Table \ref{tab:N_mod} and the
comparison with the observations is shown in Fig.~\ref{fig:rap} where we present the ratio between predicted and observed
column densities for the three molecules in the four sources for the four cases.
\begin{table}
\caption{Model column densities}
\begin{center}
\begin{tabular}{cccc}
\hline
\hline
Source & N$_{\rm SO}$ & N$_{\rm H_2S}$ & N$_{\rm CS}$ \\
\hline
MWC 480 & 6.6$\cdot 10^{12}$ & 7.9$\cdot 10^{12}$ & 2.0$\cdot 10^{13}$ \\
LkCa15 & 1.3$\cdot 10^{13}$ & 6.8$\cdot 10^{13}$ & 1.1$\cdot 10^{14}$ \\
DM Tau $^a$ & 2.2$\cdot 10^{13}$ & 6.2$\cdot 10^{13}$ & 8.0$\cdot 10^{13}$\\
\hline
MWC 480 & 7.1$\cdot 10^{11}$ & 5.5$\cdot 10^{12}$ & 1.4$\cdot 10^{12}$ \\
LkCa15 & 1.3$\cdot 10^{12}$ & 9.1$\cdot 10^{12}$ &1.1$\cdot 10^{13}$ \\
DM Tau $^a$ & 2.4$\cdot 10^{12}$ &  7.3$\cdot 10^{12}$ & 8.9$\cdot 10^{12}$ \\
\hline
\end{tabular}
\end{center}
\tablefoot{Column densities (cm$^{-2}$) of SO, CS, and H$_2$S computed by the chemical model in the four protoplanetary disks at
300~AU from the central star for a disk $5\times 10^6$~yr old. Top: Model A. Bottom: Model C.
$^a$ Modeled abundances are the same for DM Tau and GO Tau.}
\label{tab:N_mod}
\end{table}

\section{Discussion}

In Fig.~\ref{fig:rap}, except for CS, we only have upper limits on the ratios. DM Tau and GO Tau have similar model
parameters so that the chemical models predict similar results for the two disks. We chose $5\times 10^6$~yr as the
approximate age of the studied disks.

\begin{figure}
\includegraphics[width=1\linewidth]{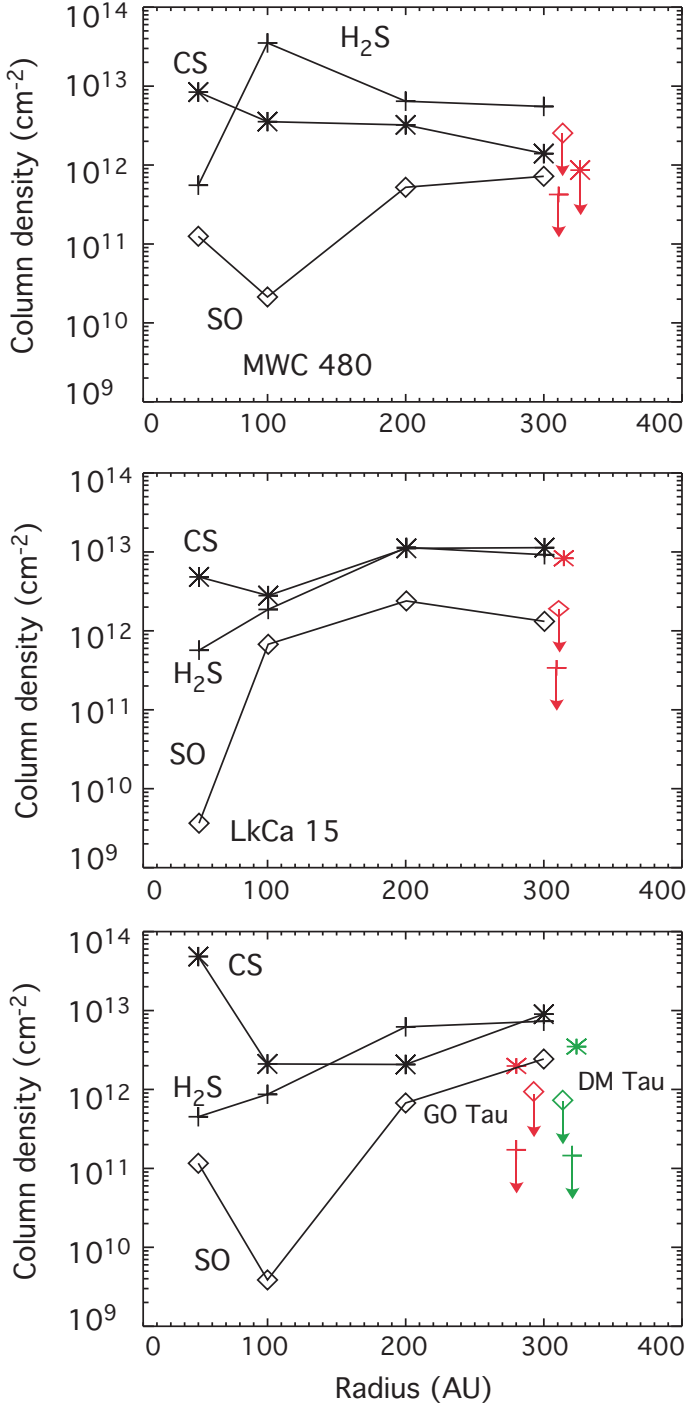}
\caption{Modeled column densities for SO (diamond), H$_2$S (cross) and CS (star) for the four sources in the case of
Model C (C/O = 1.2, initial cloud H density of $2\times 10^5$~cm$^{-3}$ and age of $10^6$~yr).
The observed column densities have been reported in grey (or color). Top panel: MWC480, medium panel: LkCa15, Bottom panel: GO Tau and
DM Tau (same model).}
\label{fig:coldens_radius}
\end{figure}

Models A and B are not very sensitive to the density of the parent cloud and its age. The difference between H$_2$S and CS
column densities are small, except for MWC 480 with the largest UV flux. When the sulfur is less abundant
by a factor of 10 (Models C and D), the results are slightly more sensitive to the density and age of the cloud.

In the case of Models C and D, corresponding to a cloud with an initial H density of $2\times 10^5$~cm$^{-3}$
and an age of $10^6$~yr, the general agreement with the observations is better. The column densities for CS
agree with the observed values, and the ones obtained for SO agree with the upper limits. The
column density for H$_2$S, in contrast, is still too high by at least one order of magnitude. This result
seems to indicate some missing processes for the destruction of H$_2$S on grains. H$_2$S is not formed efficiently
in the gas-phase \citep{Herbst_etal1989}. The gas phase H$_2$S, in these simulations, comes
from the non-thermal evaporation of solid H$_2$S formed on grains by the hydrogenation
of atomic sulfur (depleted from the gas phase). Its main destruction path in the gas-phase is through the reaction with C$^+$.
At the high densities and relatively low temperatures encountered around disk midplanes
(for H$_2$S, the evaporation temperature is around 40-50 K, \citet{Wakelam_etal2009}),
H$_2$S is likely to remain locked onto the grain surfaces and may react with other species
preventing desorption of H$_2$S. These grain surface reactions have not yet been incorporated
in chemical models. In recent experimental studies, \citet{2010A&A...509A..67G} have shown that
H$_2$S on grains is easily destroyed by cosmic-ray particles, leading to the formation of
C$_2$S, SO$_2$ and OCS on grains. In these experiments, most of the sulfur, however, may be
in the form of a sulfur-rich residuum, which could be polymers of sulfur or amorphous
aggregates of sulfur, as suggested by \citet{2004A&A...422..159W}.

We also checked the sensitivity to the C/O elemental ratio (C/O=1.2 for A and C and 0.4 for B and D).
In Models B and D with C/O=0.4, we obtain an SO column density that is higher than the
one of CS. The agreement with the observations is thus always less good than using the higher C/O value.

\begin{figure}
\includegraphics[width=1\linewidth]{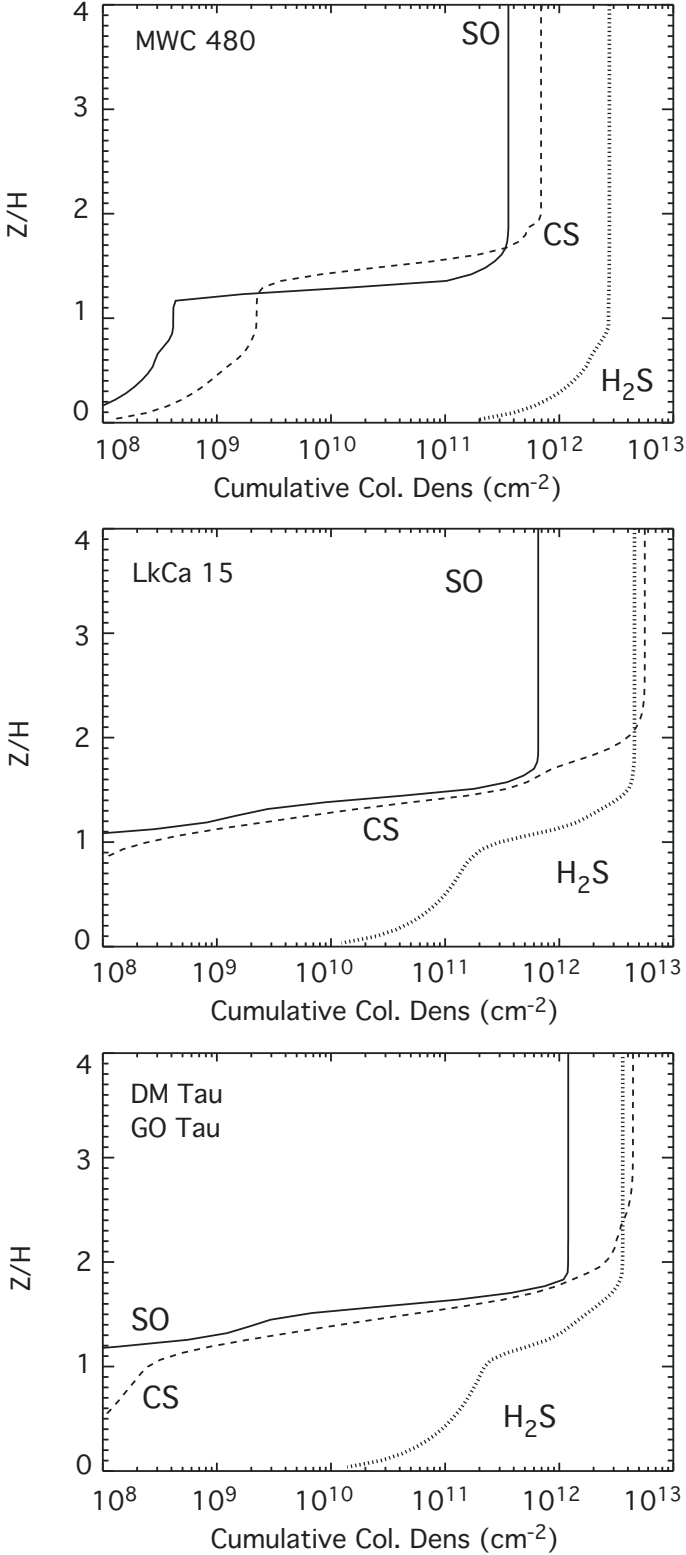}
\caption{Modeled surface densities  versus $Z/H$ for SO, H$_2$S, and CS in the case of Model C (C/O = 1.2,
initial cloud H density of $2\times 10^5$~cm$^{-3}$ and age of $10^6$~yr) at a radius of 300 AU.
Top panel: MWC480, medium panel: LkCa15, Bottom panel: GO Tau and DM Tau. Note that the cumulative column densities correspond
to half of the column density values given in Fig.\ref{fig:coldens_radius} because the vertical profiles are for half disks.}
\label{fig:cumul}
\end{figure}

We have superimposed in Fig.\ref{fig:coldens_radius} the observed column densities on the predicted column densities for the
three species in all sources for the Model C, which appears slightly better than the case of D. The column densities have been
calculated at radii, 50, 100, 200, and 300 AU. The three modeled molecular column densities show the same radial
trends in all sources with the exception of SO in MWC480 and DM Tau at 100 AU.
The difference in the SO abundances in the three sources can be explained by a balance between formation and destruction
efficiencies of this molecule. At this radius, SO is formed by the neutral-neutral reaction S + OH $\rightarrow$ SO + H and OH is
formed by the association radiation O + H $\rightarrow$ OH + h$\nu$. Water molecules are much less abundant than H or O.
It is the association between O and H that quantitatively produces OH. In MWC480 and LkCa15, the abundances of atomic O and H are
larger than in DM Tau because of larger UV fluxes, producing more SO. In MWC480, however, SO is more efficiently destroyed than
in LkCa15 because of larger abundances of ionized atomic carbon lower in the disk atmosphere (C$^+$ + SO $\rightarrow$ CO$^+$ + S).

We present in Fig.\ref{fig:cumul} the vertical surface density profiles at radius 300 AU for the sulfur-bearing
molecules for the C model, but we observe the same trends with all models. We clearly see the location of the molecular
layer at $Z/H < 2$, but the sulfur-bearing molecule abundances saturate at different heights. The SO
vertical profile is sharp and peaks at a $Z/H$ of 1.7 in DM Tau and GO Tau, 1.6 in LkCa 15 and 1.1 in MWC\,480. The H$_2$S column
densities saturate slightly below in DM Tau, GO Tau, and LkCa15. In MWC\,480, the profile is less peaked and spreads over 1.5
scale heights above the midplane. CS abundance has a rather broad maximum around two scale heights in all sources, except
MWC\,480, where it peaks at $Z/H = 1.5$ with a narrow profile.
The different behavior of MWC\,480 can be understood because the UV flux is so high that all molecules dissociate
above 1.7 -- 2 scale heights, resulting in a narrower chemically rich layer and lower overall column densities.

The relative abundances may be compared to those found in comets like Hale-Bopp \citep{Bockelee+etal2004}.
In the coma of comets, the molecules are removed from the grains by desorption. Some species, like H$_2$S, are directly desorbed
and most of them are short-lived because they are photodissociated to give daughter molecules. Both CS and SO are produced
by photodissociation of SO$_2$ and CS$_2$ \citep{Bockelee+etal2004}. Therefore, the quantity of H$_2$S observed in Hale-Bopp
is a direct tracer of the amount of H$_2$S remaining currently on cometary grains, while the CS and SO observations
most likely trace the reservoir on grains of CS$_2$ and SO$_2$.

The ratios H$_2$S/CO in Hale-Bopp and our disks are quoted in Table \ref{tab:comet}. For our disk observations,
we use the measured surface densities of CO derived from the interferometric analysis of \citet{Pietu_etal2007}
(in fact $^{13}$CO multiplied by 70, to avoid possible opacity effects). Our results are obtained at 300 AU
(the region to which we are most sensitive) and trace the gas phase, while the Hale-Bopp measurements trace
the current molecular composition of the grain mantles probably formed around 5-30 AU in the protosolar nebula.
The huge difference between observed gas ratio in the protoplanetary environment
and the cometary ice ratio, at least a factor $\sim 1000$, suggests that grain surface chemistry on
the comet may have actively modified the molecular content of the protoplanetary ice composition.

%
\begin{table}
\caption{Comparison at 300 AU between the observed ratios H$_2$S/CO in the three disks of MWC480, DM Tau
and LkCa15 and the same ratios in comet Hale-Bopp}
\begin{center}
\begin{tabular}{ccc}
\hline
\hline
Source & $\Sigma_{\mathrm{300}}$ (cm$^{-2}$)& H$_2$S/CO  \\
\hline
DM Tau  & 3.7$\times 10^{17}$ & $\leq 4 \times 10^{-5}$ \\
LkCa15 & 1.8$\times 10^{17}$ & $\leq 10^{-6}$ \\
MWC 480 & 2.3$\times 10^{17}$  &  $\leq 10^{-5}$ \\
\hline
Hale-Bopp & - & 0.07 \\
\hline
\end{tabular}
\end{center}
\tablefoot{ The CO surface densities at 300 AU \citep{Pietu_etal2007} are extrapolated from
$^{13}$CO and multiplied by 70.}
\label{tab:comet}
\end{table}

\section{Conclusions}

We have reported a new deep search for sulfur-bearing molecules (SO, H$_2$S, and CS) in three TTauri disks (DM Tau, GO Tau, and LkCa15)
and one Herbig Ae object (MWC480). CS was detected in the three TTauri disks but not in the Herbig Ae one.
H$_2$S and SO were not detected, but the upper limits are significant and allow us to make  comparisons with models. For this
purpose, we used the astro-chemistry code NAUTILUS. We find that better agreement with the observations is obtained
for an initial cloud H density of $2\times 10^5$~cm$^{-3}$ and age of $10^6$~yr with a C/O of 1.2 (as recently suggested
by \citet{Hincelin+etal_2011}).

Although it reproduces the SO and CS column densities reasonably well, our best model fails to reproduce the upper limits
obtained on H$_2$S by at least one order of magnitude, suggesting that a fraction of sulfur may be depleted in mantles or refractory grains.
At the high densities and low temperatures encountered around disk midplanes,
H$_2$S may likely remain locked onto the grain surfaces, where it may react to form other species
preventing desorption of H$_2$S. These kinds
of grain surface reactions have not yet been incorporated in models. Our results emphasize the need for grain surface
reactions in astrochemical models in presence of the high density and low temperature associated to a UV photon source,
as is the case in protoplanetary disks. More sensitive observations with ALMA, are needed to provide astrochemists
with stronger constraints on the missing ingredients in disk chemical models.

\begin{acknowledgements}
We acknowledge all the 30-m IRAM staff for their help during the observations.
This research was partially supported by PCMI, the French national program for
the Physics and Chemistry of the Interstellar Medium. We thank an anonymous referee
who provided helpful comments to improve the paper. The KIDA team is also acknowledged
for providing chemical reaction rates that are as accurate as possible for astrophysics.
\end{acknowledgements}

\bibliography{bib}
\bibliographystyle{aa}

\end{document}